\documentclass[12pt]{article}

\usepackage[T2A]{fontenc}
\usepackage[utf8]{inputenc}
\usepackage[english]{babel}
\usepackage{csquotes}

\usepackage{titlesec}
\titleformat{\section}  
  {\fontsize{14}{16}\bfseries} 
  {\thesection} 
  {1em} 
  {} 
  [] 

\titleformat{\subsection}  
  {\fontsize{10}{12}\sffamily} 
  {\S\thesubsection} 
  {2em} 
  {\color{blue}} 
  [] 

\usepackage{hyperref}
\usepackage{indentfirst}
\usepackage{latexsym}
\usepackage{float}
\usepackage{mathtools}
\usepackage{amssymb}
\usepackage{amstext}
\usepackage{amsfonts}

\usepackage{graphicx}
\usepackage[export]{adjustbox}

\usepackage{ragged2e}
\usepackage{xcolor}
\usepackage{footmisc}   
\usepackage{commath}
\usepackage{subcaption}
\usepackage{caption}
\usepackage{multirow}

\usepackage{helvet}
\usepackage{textcomp}
\usepackage{mathabx}
\usepackage{multicol}

\usepackage[font=footnotesize,labelfont=bf]{caption}
\usepackage[a4paper]{geometry}
\usepackage[
sorting=none,
doi=true,
isbn=false,
url=false,
doi=false
]{biblatex}
\renewbibmacro{in:}{}
\title{Low-mass neutron star nucleosynthesis – stripping scenario.}
\author{A. Yu. Ignatovskiy$^{1,2,*}$, I. V. Panov$^{1}$, A. V. Yudin$^{1}$}
\date{$^{1}$NIC <<Kurchatov Institute>>\\
$^{2}$Moscow Institute of Physics and Technology\\
$^{*}$Lirts@phystech.edu\\
\vspace{2ex} Date: 7 September 2024}

\newcommand{\WI}[2]{#1_{\mathrm{#2}}}

\begin{document}
\DeclareFieldFormat[article]{title}{}

\maketitle
This paper examines nucleosynthesis in a low-mass neutron star crust that loses mass due to accretion in a close binary system and, reaching a hydrodynamically unstable configuration explodes. The r-process proceeds mainly in the inner crust. Nucleosynthesis in the outer crust is an explosive process with a sharp increase in temperature caused by an outward-propagating shockwave (shock-wave nucleosynthesis). The number of heavy elements produced in a low-mass neutron star crust during the explosion is approximately $M \approx 0.041 M_{\odot}$, which exceeds the number of heavy elements ejected as jets in the neutron star merger scenario.

\noindent
\textsl{Keywords}: nucleosynthesis; nuclear astrophysics; r-process; stripping scenario.

\section{Introduction} 
\label{intro}
As a result of successful r-process modeling within the neutron star merger scenario framework, which was started more than 20 years ago \cite{Rosswog_1999_341, Freiburghaus_1999_525}, it became clear that such scenarios can play a crucial role in the heaviest nuclei formation in nature. After almost simultaneous gamma-ray burst GRB170817A observation in 2017 and the gravitational waves detection \cite{Abbott_2017_848_L12, Abbott_2017_848_L13}, as well as the lanthanides trace discovery in the accompanying kilonova \cite{Tanvir_2017_848, Domoto_2022_939, Villar_2017_851} spectrum associated with the evolution of a neutron stars close binary systems, theoretical scenarios for the r-process development \cite{Korobkin_2012_426, Cowan_2021_93, Goriely_2015_452} were confirmed.

However, the evolution of neutron stars in a close binary systems strongly depends on their masses. If the system initially has a mass asymmetry, then instead of merging a stripping scenario may take place, which, in particular, has different nucleosynthesis dynamics: \cite{Clark_1977_215, Blinnikov_1984_10, Blinnikov_2021_65, Yudin_2022_48, Yudin_2023_6} a low-mass neutron star (LMNS) is the first to fill its Roche lobe and starts to accrete onto a more massive component, becoming even lighter. As a result of such mass flow, a neutron star with a higher mass may exceed the Tolman–Oppenheimer–Volkoff limit and collapse into black hole. The low-mass component reaches the lower neutron star masses limit, which is approximately equal to $0.1M_{\odot}$ \cite{Haensel_2007_Springer}. Then, as a result of the hydrodynamic instability emergence, minimal mass neutron star (MMNS) completely collapses and explodes \cite{Colpi_1989_339, Sumiyoshi_1998_334, Blinnikov_1990_34}, enriching the interstellar medium with elements formed in a such explosion process \cite{Panov_2020_46}. 

The aim of the work presented is to obtain the nucleosynthesis results in the entire MMNS crust in the stripping scenario from instability generation up to the matter scattering to infinity, when the ejected matter can be considered as secondary ”raw material” for the next generation objects. This paper is a continuation of the studies \cite{Yudin_2023_6, Ignatovskiy_2023_86}, where the relative mass distribution of elements in the MMNS outer crust was obtained in the stripping scenario. In order to model nucleosynthesis in the inner crust it is necessary to consider the problem of matter decompression that is under sub-nuclear density conditions. The problem of MMNS core matter decompression that is initially under super-nuclear density conditions should be considered aside. Various decompression models can greatly affect the initial chemical composition one start from and, hence the final result one obtain.

The paper is structured as follows: after a brief model description and the matter state equation under consideration in section \ref{Model}, the results of nucleosynthesis for the inner and outer parts of the MMNS crust are presented, as well as integral curves for the entire crust as a whole: section \ref{Crust Nucleosynthesis}. Finally, in section \ref{Conclusions and Discussion} the stripping scenario features and its prospects are discussed.

\section{Model}
\label{Model}	
\begin{figure}[htb]
	\begin{center}
		\includegraphics[width=\textwidth]{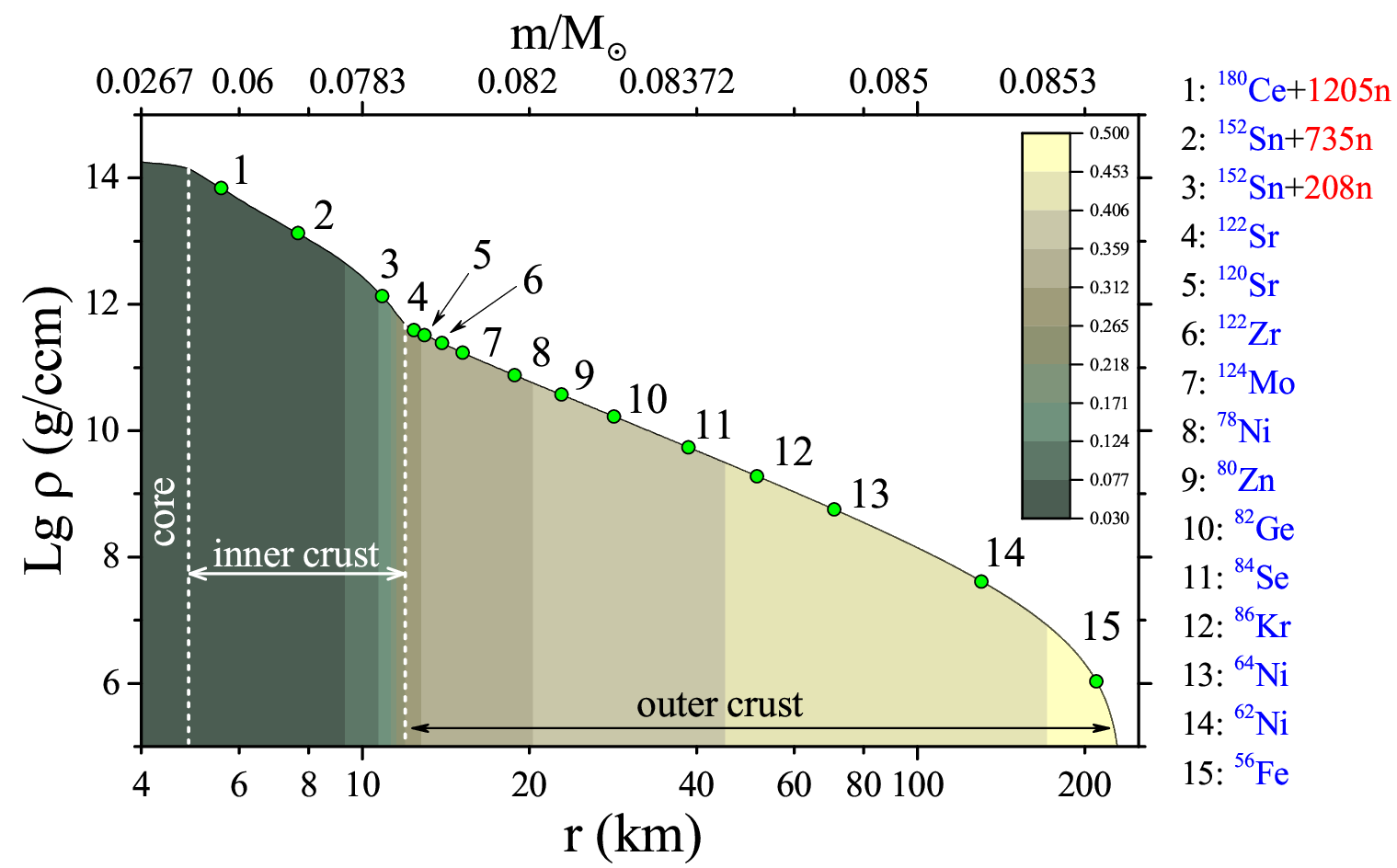}
		\caption{the MMNS structure before the explosion is shown that display density logarithm dependence on the radius. The colors indicate the electrons number per baryon $Y_{e}$ in the matter. The inner and outer crust are divided into layers 1-15, initial chemical composition and numbering of which are given to the figure right.}
		\label{figure:2}
	\end{center}
\end{figure}

Nucleosynthesis was modeled based on calculations performed by Ref. \cite{Yudin_2022_48} that use  relativistic hydrodynamics equations~\cite{Hwang_2016_833}. Constructed on the Skyrme functional BSk25~\cite{Pearson_2018_481} was chosen as the equation of state (EoS) that corresponds to the observational constraints on the neutron stars masses. The dependence of nucleosynthesis results on the equation of state was previously studied in Ref. \cite{Ignatovskiy_2023_86}. Within the framework of the cold catalyzed matter approximation, there is an energetically most favorable state \cite{Haensel_2007_Springer} for a given density value -- the MMNS crust is subdivided into layers that consist one type nucleus surrounded by electrons in the outer crust or immersed in a free neutrons <<sea>> and electrons in the inner crust (see Fig. \ref{figure:2}). Each layer is characterized by a trajectory -- the dependence of substance density and temperature on time from the moment the instability generates. Along each trajectory (with the initial chemical composition of the corresponding layer) nucleosynthesis is modeled -- a system of a large number of stiff non-linear differential equations is calculated numerically under variable environmental conditions based on the modified SYNTHER code \cite{Korneev_2011_37}, which implements the kinetic implicit scheme \cite{Panov_2001_27, Blinnikov_1994_266, Gear_1971}. The jacobian coefficients are the rates of all possible nuclear reactions: pair reactions with neutrons, protons, alpha particles and gamma quanta~\cite{Rauscher_2000_75}; beta decays~\cite{Moller_1997_66} and their inverse reactions of electron and positron capture~\cite{Langanke_2000_673}; beta-delayed neutron emissions~\cite{Moller_2003_67}; spontaneous, induced and beta-delayed fission~\cite{Korneev_2011_37, Panov_2005_747, Panov_2010_513, Panov_2013_39}. Experimental beta decay rates were taken from the nuclear database NuDat2 (2009)\footnote{http://www.nndc.bnl.gov/nudat2}.

The nuclei involved region in the simulation was determined by the extended Thomas-Fermi mass model with the Strutinskii integral correction (ETFSI)~\cite{Aboussir_1995_61}: $Z=0-110$, $A^{\mathrm{max}}(Z)$ is the last heaviest isotope with a positive neutron binding energy $\WI{S}{n}>0$ for a given atomic charge number Z. The total number of nuclei (as well as the stiff non-linear differential equations number) is 5802.

From Fig. \ref{figure:2} one can make certain that the MMNS radius exceeds 200 km, which is very different from the regular 10 km often associated with this objects type. As mass decreases, the neutron stars radius sharply increases~\cite{Lattimer_2004_304, Yudin_2022_48}. It is worth noting that single neutron stars cannot reach such size due to the conditions absence for mass loss. The stripping scenario is possibly the only source of such objects birth.

The inner crust matter (layers 1-3) is initially at a sub-nuclear density and consists of neutron-rich exotic nuclei immersed in a <<sea>> of free neutrons~\cite{Haensel_2007_Springer}. The question of how these clusters evolve during the matter decompression that occurs during the MMNS explosion is still open. In this paper, we assume that as the density decreases, these clusters undergo multiple neutron evaporation through the $(\gamma,n)$ channel, rapidly approaching the nuclear stability boundary. Thus, the initial chemical composition transforms into nuclei with approximately nullified neutron binding energy $\WI{S}{n} \approx 0$, free neutrons and electrons.

In Ref. \cite{Yudin_2023_6}, the impact of two different limiting decompression channels on nucleosynthesis in the stripping scenario was investigated: evaporation $(\gamma,n)$ of neutrons and $\beta^{-}$ decays. The results demonstrated a weak dependence of the final abundance on the decompression type, which is probably due to the strong nuclear fission cycling influence \cite{Panov_2005_747} for the strong r-process occurring at a free neutron to seed ratio of $\WI{N}{n}/\WI{N}{seed} \gtrsim 150$ \cite{Meyer_1992_399, Thompson_2001_562, Panov_2009_494} (this condition is satisfied for the entire inner crust).

\begin{figure}[H]
	\begin{center}
		\begin{subfigure}[b]{0.495\textwidth}
			\centering
			\includegraphics[width=\textwidth]{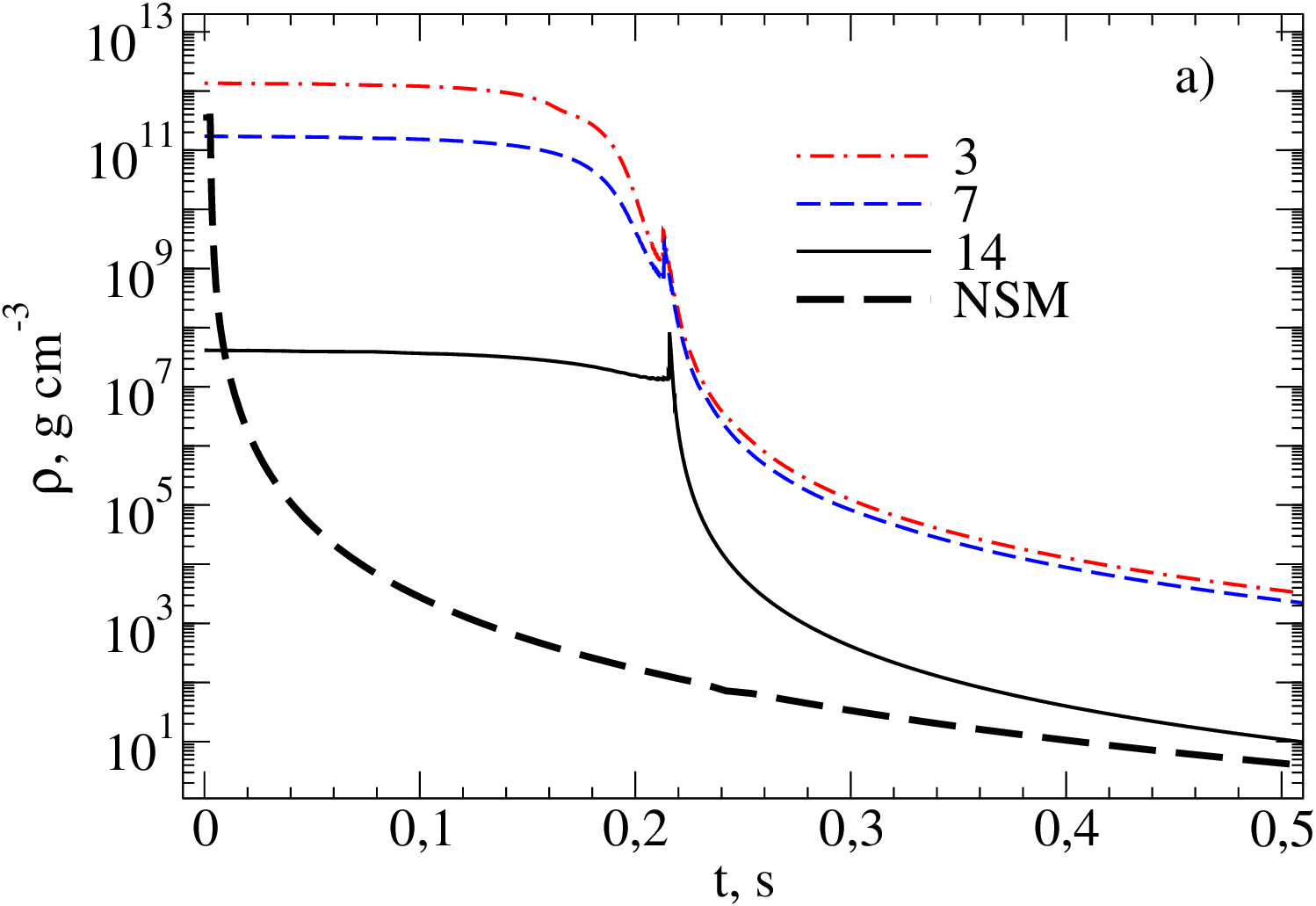}
		\end{subfigure}
		\hfill
		\begin{subfigure}[b]{0.495\textwidth}
			\centering
			\includegraphics[width=\textwidth]{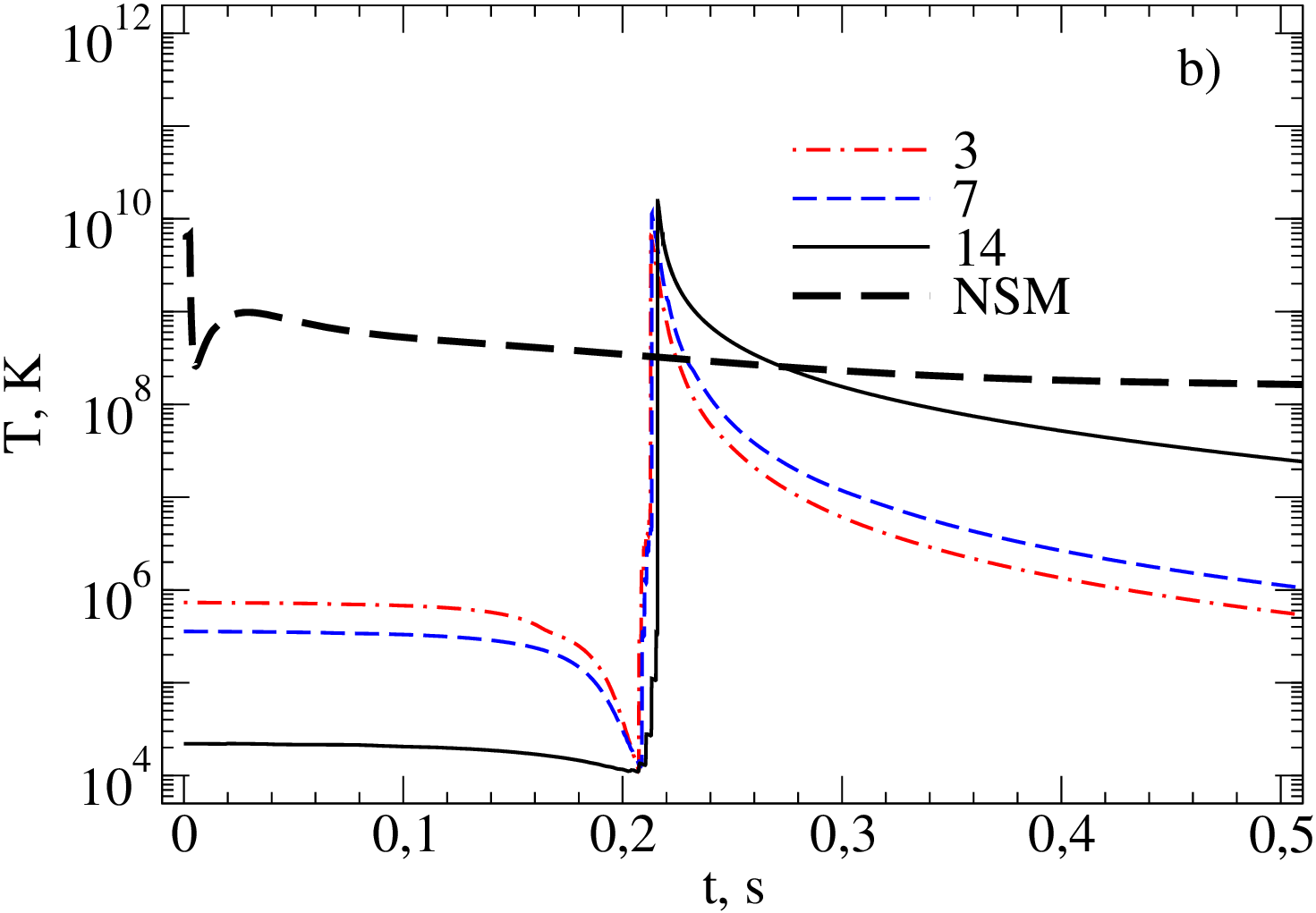}
		\end{subfigure}
		\caption{expanding matter density (a) and temperature (b) dependence on time in the stripping scenario for trajectories № 3, 7 and 14 are shown, as well as evolution curves for NSM of approximately equal masses.}
		\label{figure:3}	
	\end{center}
\end{figure}

Fig. \ref{figure:3} shows some of the most representative inner (a) and outer (b) crust trajectories, as well as a typical trajectory of the neutron star merger scenario (NSM presented as dashed line). The matter expansion in the stripping model occurs slower than the jet emission in the NSM scenario, which leads to different dynamics of density change with time in these scenarios. A distinctive feature of the stripping scenario is the matter strong heating caused by the shock wave arrival at the time $t_{sw} \approx 213$ ms for the EoS considered. The heating-up mentioned leads to a sharp speed-up of direct and, for heavy nuclei, inverse reactions rates, as well as photo-nuclear reactions rates, leading to explosive nucleosynthesis that we call here <<shock-wave  nucleosynthesis>> to emphasize its short and rapid character.

\section{Crust Nucleosynthesis}
\label{Crust Nucleosynthesis}

\begin{figure}[H]
	\begin{center}
		\begin{subfigure}[b]{0.495\textwidth}
			\centering
			\includegraphics[width=\textwidth]{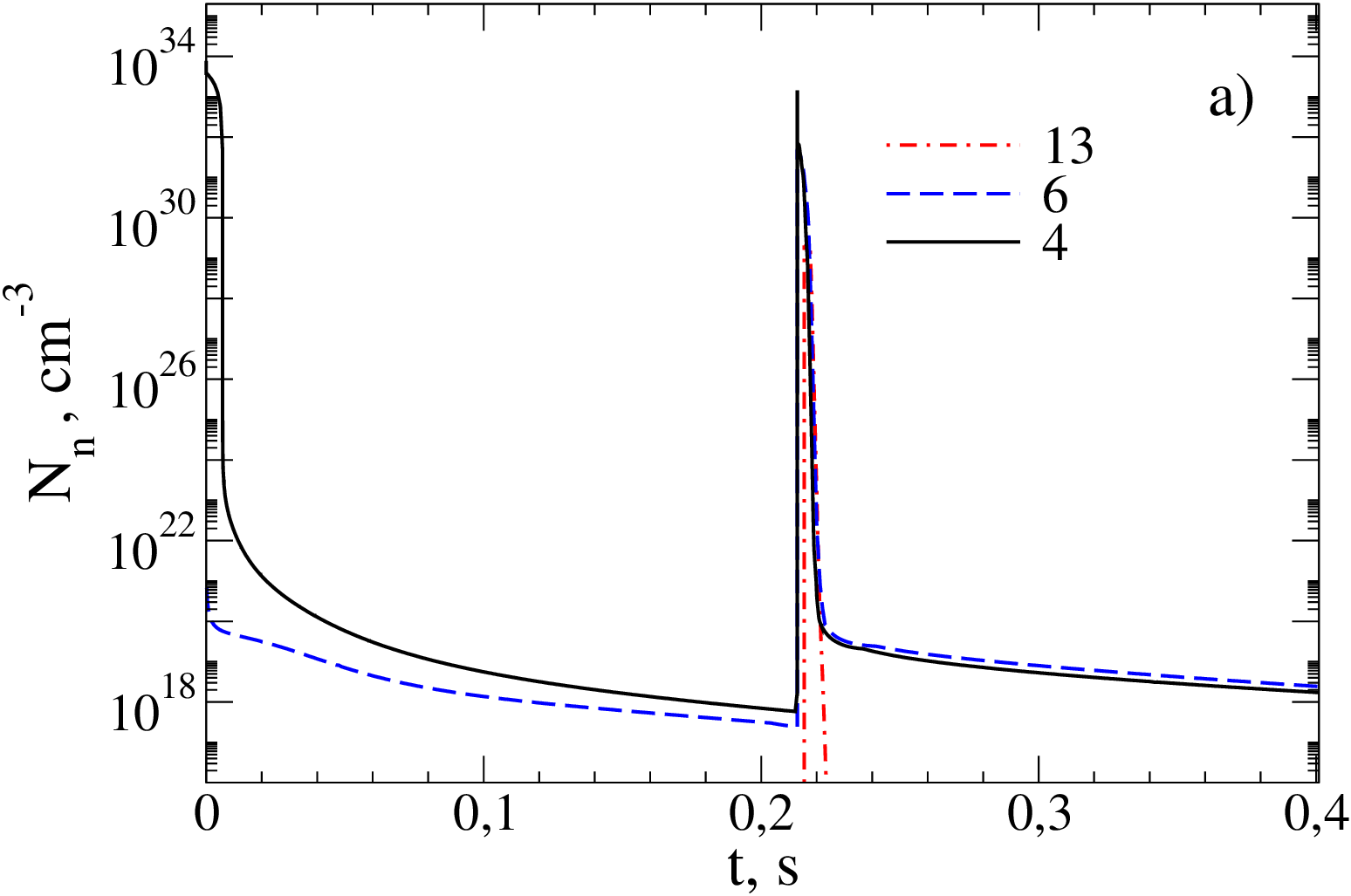}
		\end{subfigure}
		\begin{subfigure}[b]{0.495\textwidth}
			\centering
			\includegraphics[width=\textwidth]{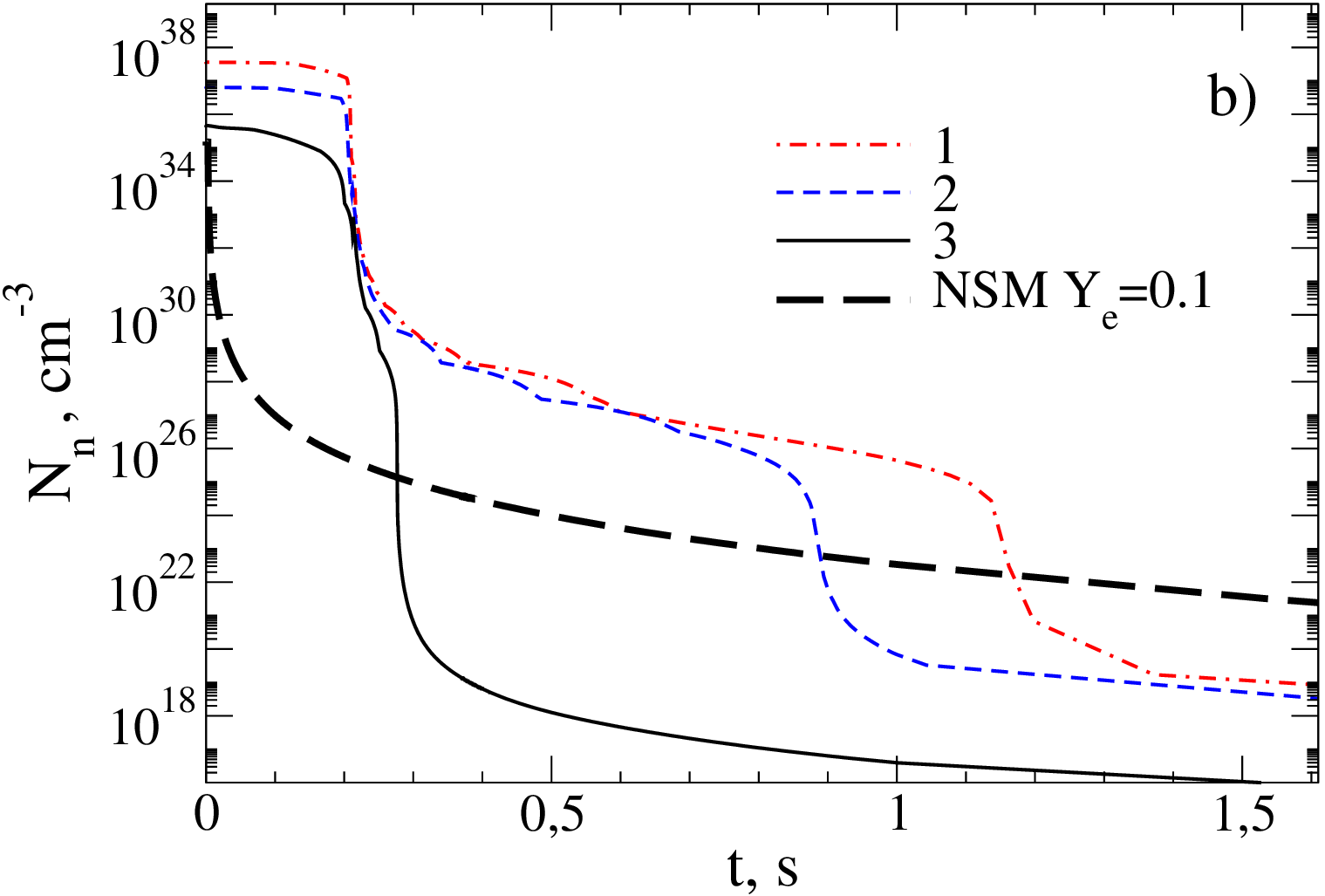}
		\end{subfigure}
		\caption{free neutron concentration $\WI{N}{n}$ dependence on time $t$ for characteristic trajectories. (a): outer crust. (b): inner crust and the NSM trajectory with initial $Y_e=0.10$ (b). The curve number is the stripping scenario trajectory number (see Fig. \ref{figure:2}).}
		\label{figure:4}
	\end{center}
\end{figure}

Fig. \ref{figure:4} shows the free neutrons concentration dynamics for the outer (a) and inner (b) MMNS crust trajectories. During the transition to the shock-wave nucleosynthesis in the outer crust, at the moment of the shock wave arrival, a neutrons fraction is released from the nuclei due to the strong increasing of reverse reactions rates (Fig. \ref{figure:4}a). The jump is not noticeable in the inner crust (Fig. \ref{figure:4}b), which is associated, firstly, with weaker heating compared to the outer crust, and secondly, the inner crust substance does not have time to exhaust all the free neutrons reserves available for the r-process before the shock wave arrives.
\\

In the outer crust (Fig. \ref{figure:4}a), the initial free neutrons concentrations either do not correspond to the r-process requirements $\WI{N}{n}<10^{22}$ cm$^{-3}$ (for example, trajectories № 6 and 13), or correspond, but are maintained at the required values for an insufficient amount of time $\Delta t \approx$ 1-10 ms (trajectory № 4), which is comparable with the characteristic beta-decays $\tau_{\beta}$ half-lives. Nucleosynthesis in the outer crust is represented mainly by a n-process and explosive nucleosynthesis composition. A strong r-process occurs throughout the inner crust.

\begin{figure}[H]
	\begin{center}
		\begin{subfigure}[b]{0.495\textwidth}
			\centering
			\includegraphics[width=\textwidth]{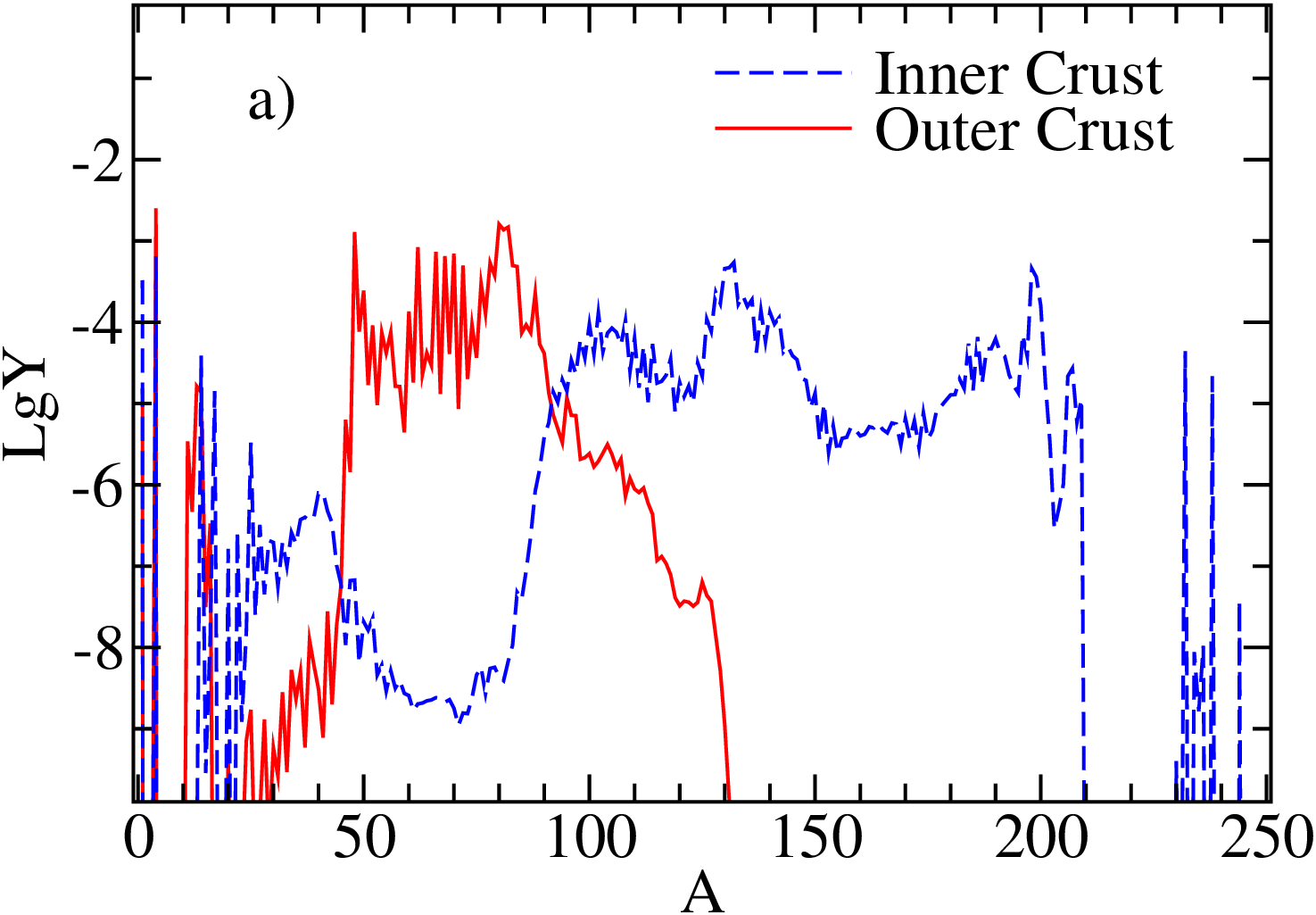}
		\end{subfigure}
		\begin{subfigure}[b]{0.495\textwidth}
			\centering
			\includegraphics[width=\textwidth]{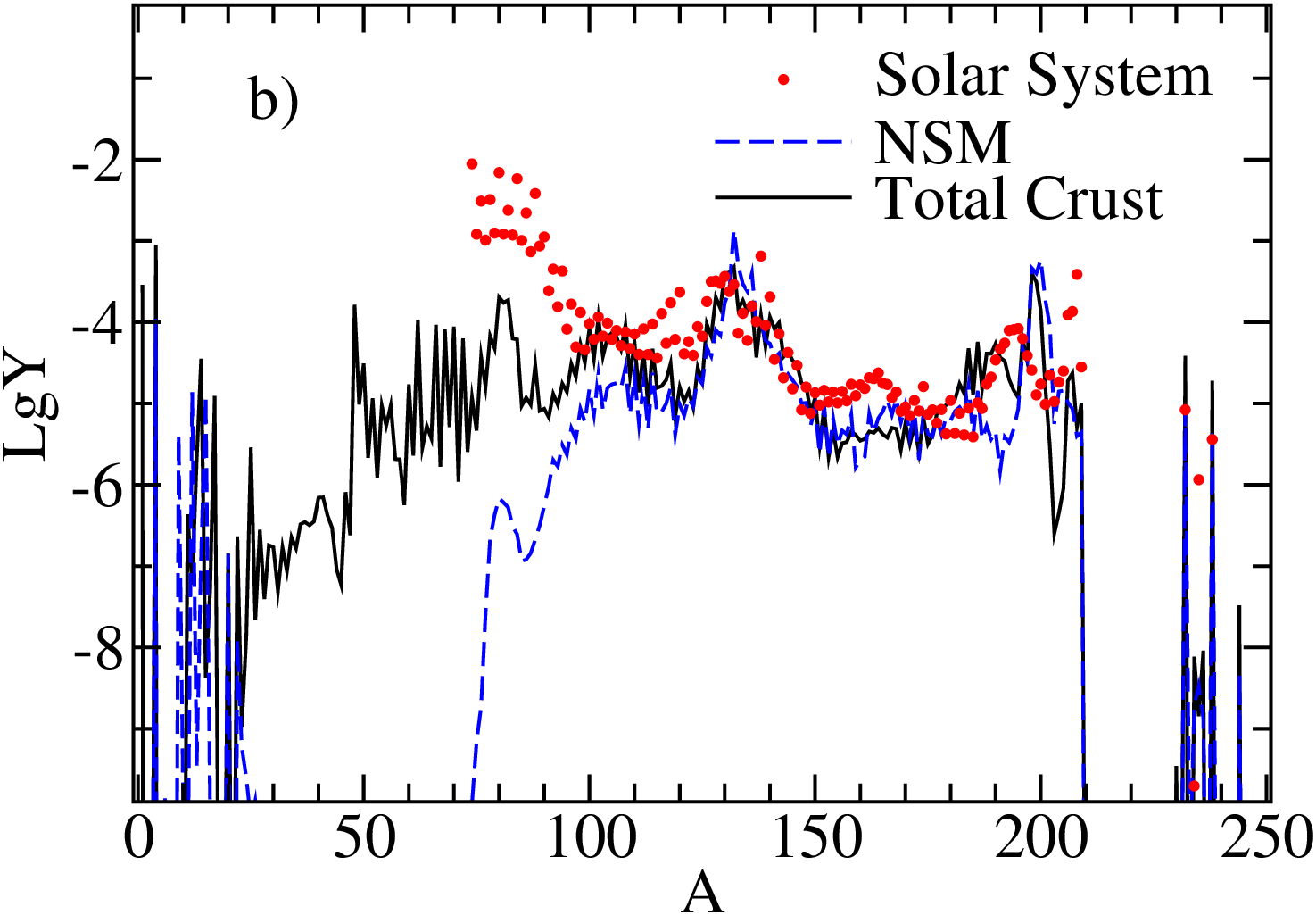}
		\end{subfigure}
		\caption{reduced concentration logarithm  $\lg Y$ on atomic mass number $A$. (a): outer crust (solid red) and inner crust (dashed blue) integral curves in the stripping scenario. (b): integral curve for the entire crust (solid black) in the stripping scenario, averaged over $\WI{Y}{e}$ integral curve in the NSM scenario \cite{Rosswog_1999_341} (dashed blue), and the observed solar abundance \cite{Lodders_2019} (red markers). The normalization for the stripping and NSM curves is $\sum A_{i}Y_{i}=1$. For the solar system abundance, the normalization is chosen arbitrarily to compare the relative different element groups contributions.}
		\label{figure:6}
	\end{center}
\end{figure}

Fig. \ref{figure:6}a shows the nucleosynthesis integral $\lg Y(A)$ ($Y=nm_{\mu}/\rho$) curves for the outer (solid red) and inner (dashed blue) crust in the stripping scenario. Fig. \ref{figure:6}b shows the nucleosynthesis integral $\lg Y(A)$ curves for the entire crust (solid black) in the stripping scenario, mean $\WI{Y}{e}$ curve for the NSM scenario (dashed blue) and the observed solar abundance~\cite{Lodders_2019} (red markers). The normalizations for the stripping and merger scenarios are chosen such that $\sum A_{i}Y_{i}=1$. For the solar abundance, the normalization is chosen arbitrarily and is plotted to compare the relative different element groups contributions. Note that in the stripping scenario more light elements $A=10-80$ are synthesized, with the $A=50-80$ region being formed predominantly in the outer crust (see Fig. \ref{figure:6}a), and the $A=10-50$ region being formed in the inner crust (see Fig. \ref{figure:6}a).
\\
\\

\section{Conclusions and Discussion}
\label{Conclusions and Discussion}

In this paper, we studied inner and outer crust nucleosynthesis during the MMNS explosion in the stripping scenario. The layers initial composition varied greatly in both chemical elements and neutron excess (see Fig. \ref{figure:2}). The initial value of $\WI{Y}{e}$ varied from 0.042 for trajectory № 1 (inner crust) up to 0.452 for trajectory № 15 (outer crust).
We demonstrated that with such a wide range of parameters, various types of nucleosynthesis are possible, caused by short but strong substance heating by the shock wave.

To order to account for the sub-nuclear density matter decompression, an assumption of $(\gamma,n)$ channel neutron evaporation was used in which exotic nuclei lose neutrons via photodissociation reactions until they pass into the positive neutron binding energies region $\WI{S}{n}>0$. In reality, exotic nuclei can pass into the $\WI{S}{n}>0$ region in a somewhat more complex way that consist of a different reaction channels combination with different weight fractions, for example, $\beta$ decays, neutrons photodissociations $(\gamma, n)$ and nuclear fission. Earlier~\cite{Yudin_2023_6} we found that different limiting decompression channels have a weak effect on the chemical elements final abundance due to nuclear fission cycling, which is carried out for all MMNS inner crust trajectories. These results are encouraging for the future study of MMNS core nucleosynthesis, where the neutron excess is higher compared to the inner crust.

The stripping scenario for nucleosynthesis is interesting for at least two reasons: 1) the flare energy is about $10^{47}$~erg and the GRB170817A event ejection sphericity are better described within the stripping model framework~\cite{Blinnikov_2022_5}; 2) the total mass of the ejected matter in the stripping scenario is equal to the entire MMNS mass at instability generation moment $\approx 0.1 M_{\odot}$, which is several times greater than the matter ejected  mass as a jet in the NSM scenario (see, for example, Ref. \cite{Korobkin_2012_426}).

Although the 2017 event observational characteristic rather point in the stripping scenario favor, which better describes some solar abundance regions, the elemnts integral abundance in the Universe is a all possible nucleosynthesis scenarios composition, some of which may not yet be discovered, and the others relative event frequencies are unknown.

The heavy ($A \gtrsim 60$) elements mass fraction ejected into the interstellar medium in the stripping scenario is $\approx 97\%$, and the MMNS crust mass alone is $\approx 0.04 M_{\odot}$, which is greater than the entire ejected matter mass in the NSM scenario, which consists almost entirely of heavy elements $\approx 100\%$. Thus, if we assume that both of these scenarios are realized in nature with approximately the same frequency~\cite{Kramarev_2023_525}, the stripping scenario will provide a larger total contribution to the chemical elements existing abundance.

Strong outer crust heating accelerates not only photoreactions, but also the electron and positron captures by nuclei rates, which are presented in this code as the iron-peak data~\cite{Langanke_2000_673} extrapolations. The lepton capture rates influence for elements heavier than iron-peak on the final chemical elements abundances under explosive nucleosynthesis conditions other than supernova explosions will be described in details in a separate paper.

\section*{Acknowledgements}

This work was done within the National Research Center <<Kurchatov Institute>> state assignment framework. Nucleosynthesis calculations performed by A. Yu. Ignatovsky were supported by the BASIS Foundation, grant No. 24-1-5-66-1.

\printbibliography

\end{document}